\documentclass[a4paper,10pt,twoside]{cpc-hepnp}
\usepackage{multicol}
\usepackage{multirow}
\usepackage{graphicx}
\usepackage{booktabs}
\usepackage{amssymb,bm,mathrsfs,bbm,amscd}
\usepackage[tbtags]{amsmath}
\usepackage{lastpage}
\usepackage{CJK}

\begin{document}

\begin{CJK*}{GBK}{song}

\fancyhead[c]{\small Submitted to Chinese Physics C} \fancyfoot[C]{\small 
\thepage}
\footnotetext[0]{Received XXX}

\title{The calibration of DD neutron indium activation diagnostic for Shenguang-\uppercase\expandafter{\romannumeral3} facility$^{*}$}

\author{
      SONG Zi-Feng
\quad CHEN Jia-Bin
\quad LIU Zhong-Jie
\quad ZHAN Xia-Yu
\quad TANG Qi$^{1)}$\email{Corresponding author: E-mail: tangqilfrc@caep.cn}
}
\maketitle

\address{
$^{}$ Research Center of Laser Fusion, China Academy of Engineering Physics, Mianyang 621900, China\\
}

\begin{abstract}
The indium activation diagnostic was calibrated on an accelerator neutron source in order to diagnose deuterium-deuterium (DD) neutron yields of implosion experiments on Shenguang-III facility. The scattered neutron background of the accelerator room was measured by placing a polypropylene shield in front of indium sample, in order to correct the calibrated factor of this activation diagnostic. The proper size of this shield was given by Monte Carlo simulation software. The affect from some other activated nuclei on the calibration was verified by judging whether the measured curve obeys exponential decay and contrasting the half life of the activated sample. The calibration results showed that the linear range reached up to 100 cps net count rate in the full energy peak of interest, the scattered neutron background of accelerator room was about 9\% of the total neutrons and the possible interferences mixed scarcely in the sample. Subtracting the portion induced by neutron background, the calibrated factor of our sample was $4.52\times10^{-7}$ counts/n with uncertainty of 4.3\%.
\end{abstract}



\footnotetext[0]{\hspace*{-3mm}\raisebox{0.3ex}{$\scriptstyle\copyright$}2014
Chinese Physical Society and the Institute of High Energy Physics
of the Chinese Academy of Sciences and the Institute
of Modern Physics of the Chinese Academy of Sciences and IOP Publishing Ltd}%

\begin{multicols}{2}

\section{Introduction}
The experimental ignition threshold factor (ITFx) is a performance criterion that indicates the quantitative implosion performance, and it is used to judge the proximity of the ignition threshold for inertial confinement fusion (ICF) experiment \cite{lab1,lab2,lab3,lab4}. ITFx is related to two independent variables which are neutron yield and total areal density. Thus the neutron yield is one of the most important parameter to characterize ICF experiment performance. The activation diagnostic is a good substitute for the plastic scintillator detector at higher neutron yield ( $\textgreater 10^{10}$) \cite{lab5,lab6,lab7,lab8}, which will improve the neutron yield measurement accuracy of implosion experiments conducted on Shenguang-\uppercase\expandafter{\romannumeral3} laser facility. Currently, the implosion capsules used on Shenguang-\uppercase\expandafter{\romannumeral3} facility are mostly filled with deuterium-deuterium (DD) fuel, and the DD neutron yield can be measured by an indium activation diagnostic \cite{lab9,lab10,lab11}. The neutron yield measurement accuracy by this activation diagnostic plays an important role in the estimation of implosion performance and the evaluation of capsule design.

To measure DD neutron yield accurately, this activation diagnostic needs calibrating precisely before experiments. A factor can be calibrated on an accelerator neutron source to obtain the indium sample activity induced by incident neutron.  Gary. W. Cooper \cite{lab10} and Olivier Landoas \cite{lab11} had many previous work of indium activation calibration on different neutron sources, and they got the uncertainties of calibrated factor were 8\% and 7.5\%, respectively. During calibration, the scattered neutron background from the accelerator room will enter the sample and increase the activated atoms in the indium sample. At the same time, the possible interference nuclides in the sample will have a more or less influence on the calibration factor. Thus in this paper, the scattered neutron background from the accelerator room was measured by placing a neutron shield in front of the indium sample in order to correct the calibrated result. And the influence of possible interference nuclides on the calibration factor was estimated by the decay curve measurement technique. Furthermore, the linear range of this activation diagnostic was calibrated by the associated particle method.

\section{The measurement principle}
Among all the materials possibly available \cite{lab5,lab12}, indium is the most proper for DD neutron measurement. The indium sample is composed of two isotopes $^{115}$In and $^{113}$In, and their natural abundances are 95.7\% and 4.3\%, respectively. It can be used to measure the 2.45 MeV neutrons via the neutron inelastic scattering reaction with $^{115}$In: $^{115}$In(n, n$^{'}$)$^{115m}$In \cite{lab10,lab11}. This reaction has a lower threshold of 336 keV, and so it leads to a slight sensitivity to the possible scattered neutrons. The metastable state of the reaction product $^{115m}$In has a half life of 4.486 h and emits a 336.24 keV gamma ray with an intensity of 45.9\%.

The activated indium sample is transported to a lead shielded high-purity germanium detector (HPGe) containing a ORTEC digital gamma ray spectrometer whose model is DSPEC-jr-2.0, in order to perform radiation measurement for the gamma ray of interest. When an indium sample with known size is exposed in a neutron source, the net count (\emph{$N_{\gamma}$}) of 336.24 keV gamma ray detected by HPGe is directly proportional to the neutron yield (\emph{$Y_{n}$}), as follow
\begin{equation}
\label{eq1}
N_\gamma=Y_nF0.5^{t_c/T}(1-0.5^{t_m/T}),
\end{equation}
where \emph{F} is the sensitivity of indium activation diagnostic with meaning of the average activity of indium sample induced by per neutron, $t_{c}$ and $t_{m}$ are cooling time and measuring time for gamma ray measurement, \emph{T} is the half life of radioactive production. The \emph{F} factor is mainly related to the self absorption factor of gamma in the sample, the detection efficiency to the gamma ray of interest and the solid angle of sample to neutron source. The self absorption factor depends on the thickness of sample, and the proper thickness is 1 cm according to the simulation result by Monte Carlo codes, as shown in Fig.~\ref{fig1}.
\begin{center}
\includegraphics[width=7cm]{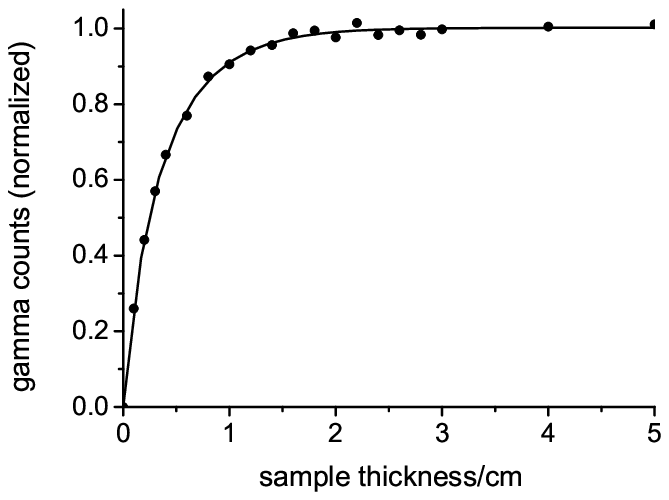}
\figcaption{\label{fig1} The normalized counts of gamma rays vary with sample thickness. The dots are normalized gamma counts escaping from the irradiated front side of indium sample which are simulated by Geant4.9.2.p04 codes, and the solid line is a fitted curve based on these dots by least square method. The escaping gamma will approach the maximum when the sample thickness is 1 cm.}
\end{center}

\section{The setup of calibration experiment}
The appropriately sized indium sample is exposed to a neutron source with known flux \cite{lab10,lab11}. Then we can get the \emph{F} factor by taking the measured gamma counts, the cooling time and the measuring time into Eq. (1). Knowing the indium activity induced by incident neutrons, we can measure unknown neutron yields from implosion capsules.

The indium activation diagnostic was calibrated on the accelerator neutron source located at Institute of Nuclear Physics and Chemistry with associated particle method. To produce 2.45 MeV neutrons, deuterium particles beam with a average energy of 140 keV impacts a titanium-deuterium (TiD) target using D(d, n)$^{3}$He reaction, which brings about neutrons between 2.1 MeV and 2.9 MeV depending on detection direction. The accelerator neutron flux is deduced from the protons, which come from D(d, p)T reaction and are measured by a gold-silicon surface barrier detector in the direction of 135$^{\circ}$ relative to the deuteron beam. A 2-$\mu$m-thick aluminum foil was placed in front of the detector to reject the background due to elastically scattered deuterons from the target. A 7 cm diameter by 1 cm thick indium sample was placed at 95$^{\circ}$ at a distance of 40 cm from the target. The neutrons entering the front surface of indium sample had energy of 2.45$\pm$0.04 MeV. After irradiation, the activated indium sample was transported to the HPGe detector for its radiation decay measurement. Fig.~\ref{fig2} is the schematic of the experimental setup. 
\begin{center}
\includegraphics[width=4.5cm]{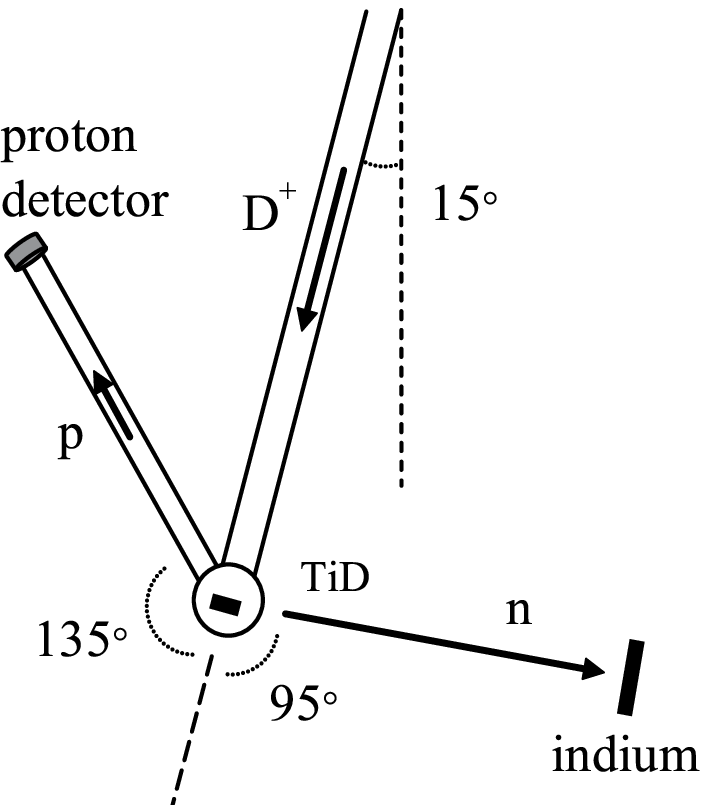}
\figcaption{\label{fig2} The schematic of the experimental setup.}
\end{center}

The number of activated atoms ($N^{*}$) in the sample is directly proportional to the neutron numbers entering the sample, and it is given by
\begin{equation}
\label{eq2}
N^*=\chi \Omega Y_i,
\end{equation}
where \emph{$\chi$} represents the number of activated atoms induced by per neutron, $\Omega$ is the solid angle of sample and $Y_{i}$ is the neutron yield of accelerator source. In order to obtain a sufficient activity of sample, the indium sample is usually exposed to the neutron source for a long-time irradiation. However, the neutron flux during long irradiation is not a constant owing to instable deuterium beam, and the activated atoms decay during irradiation stage. The neutron yield is deduced by the proton numbers detected by surface barrier detector. During irradiation stage, the proton numbers are counted respectively in many short time intervals ($\Delta$t). Supposing that the proton count of the \emph{i}-th interval is $N_{i}$, the residual activated atoms (\emph{$N_{\gamma}^{*}$}) at the end of irradiation which are produced in this interval are give by
\begin{equation*}
N_\gamma^*=\chi \Omega N_iS_p0.5^{(t_r-i\Delta t)/T},
\end{equation*}
where $t_r$ is the total irradiation time and $S_p$ is the sensitivity of proton detector which represents neutron numbers of per proton detected by surface barrier detector. Summing all intervals, the accumulating residual atoms (\emph{$N_c^*$}) in the indium sample are given by
\begin{equation*}
N_c^*=\sum_{i=1}^n N_\gamma ^*,
\end{equation*}
with $n=t_r/{\Delta}t$. The activity of residual activated atoms is equivalent to that induced by transient irradiation with the yield $Y_c$, according to Eq. (2), then we can get
\begin{equation}
\label{eq3}
Y_c=0.5^{t_r/T}S_p\sum_{i=1}^n N_i 0.5^{-i\Delta t/T}.
\end{equation}

\section{The result of calibration experiment}

The activity of a sample is directly proportional to the neutrons entering the sample. If the sample is at a too high activity, the activation diagnostic will miss detecting a part of radioactive rays because of the dead-time of electronics. The linear range of the activation diagnostic wants measuring to confirm the maximum radioactive activity applied to this diagnostic. An indium sample was exposed to an accelerator neutron source for sufficient irradiation, and then this activated sample was transported to the HPGe detector. The gammas of interest were registered for 48 hours in one hour interval by the spectrum software GammaVision-32 with a JOB command. The maximum dead-time on the germanium detector was lower than 10\%. From the gamma counts of per one hour, the \emph{F} factors in each time period can be obtained by using Eq. (1) with a dead-time correction technique. The \emph{F} factors will be constant without count saturation. Fig.~\ref{fig3} is the measured results of \emph{F} factor for each one hour. The errors of the factor in lower count rates mainly come from the larger statistical fluctuation because of fewer detected gamma rays. It can be seen that the factors are consistent within their uncertainties, and the linear range of this diagnostic reaches up to about 100 cps (counts per second) within the full energy peak. When a 7 cm diameter by 1 cm thick sample is placed at a distance of 40 cm from target, the maximum neutron yield is about $5\times10^{12}$ that is deduced by the activity of sample. If the neutron yield becomes higher for future implosion experiment, the dead-time of the germanium detector also can be controlled within 10\% by using a smaller size sample or a farther distance.
\begin{center}
\includegraphics[width=7cm]{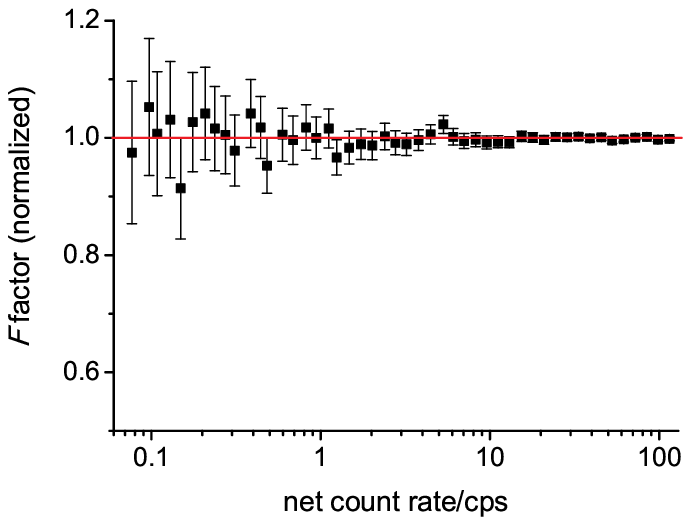}
\figcaption{\label{fig3}  (Color online) The measured results of \emph{F} factor in different net count rate within full energy peak of interest. The squares are the measured results of \emph{F} factor with uncertainties of k=2 and the red line is the average factor used as the normalized value.}
\end{center}

There are some nuclides, such as $^{59}$Fe (4.6 s), $^{72}$Ga (46.5 h), $^{88}$Kr (16.5 s), $^{89}$Kr (4.4 s), $^{101}$Mo (7.1 s) and so on, also emit gamma rays with energy near 336 keV resolved hardly by the germanium detector, excluding $^{115m}$In (4.486 h). A decay curve measurement technique is used to check for these possible interferences by contrasting the half life of activated atoms. The decay curve of activated sample is according to that of $^{115m}$In nuclide within twice of the statistical fluctuation from Fig.~\ref{fig4}. And it indicates that the possible interferences mix scarcely in the sample.
\begin{center}
\includegraphics[width=7cm]{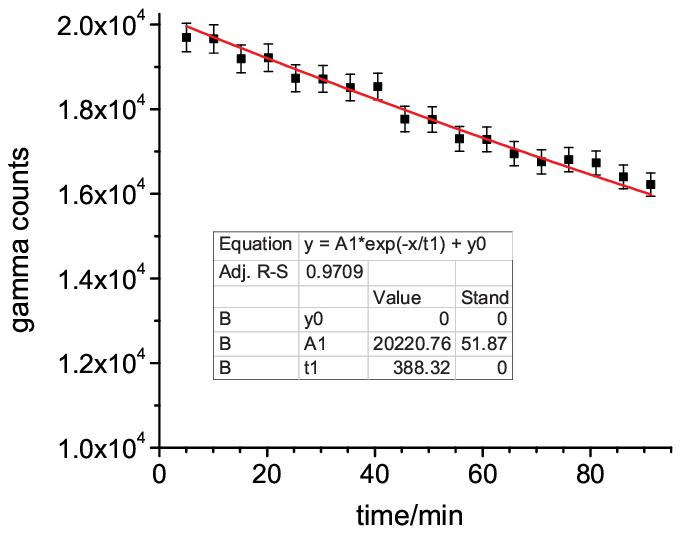}
\figcaption{\label{fig4}  (Color online) The measured decay curve of the activated sample during activation calibration. The squares are the gamma counts measured in per 5 minutes with uncertainties of k=2. The red solid line is the decay curve of $^{115m}$In with $1/\lambda$=388.32 min (marked as t1 in the equation).}
\end{center}

The indium samples were exploded to the neutron source for a 10 minutes irradiation with proton counted in per 100 seconds, and the neutron yields at the end of irradiation were about $1\times10^{11}$ with an uncertainty lower than 0.2\% from statistical fluctuation. After irradiation, the samples were transported to a coaxial high-purity germanium detector for gamma spectrum measurement. The typical gamma spectrum measured during calibration is shown in Fig.~\ref{fig5}. 
\begin{center}
\includegraphics[width=7cm]{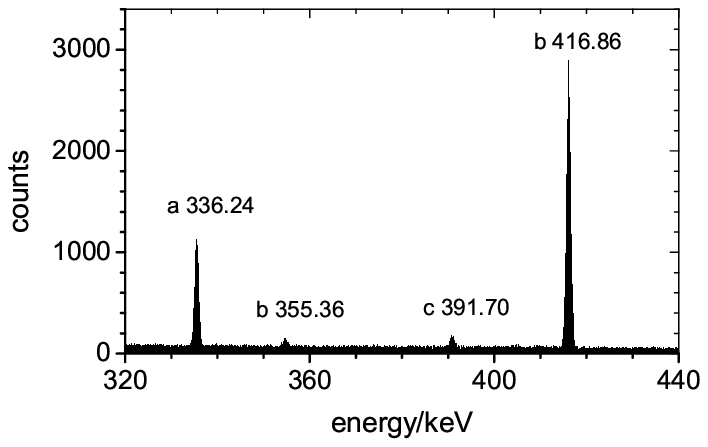}
\figcaption{\label{fig5}  The typical gamma spectrum measured during calibration. The main full-energy peaks come from the isotope $^{115}$In or $^{113}$In reacting with neutrons: (a) $^{115}$In(n, n')$^{115m}$In, (b) $^{115}$In(n, $\gamma$)$^{116}$In, (c) $^{113}$In(n, n')$^{113m}$In. The reaction (a) and (c) are sensitive to 2.45 MeV DD neutrons, meanwhile reaction (b) is more sensitive to lower-energy neutrons than DD neutrons.}
\end{center}

The sample was cooled 10 minutes after irradiation, and then transported to the HPGe detector to measure the gamma ray of interest for 1.5 hours. The \emph{F} factor of each sample was calculated by using Eq. (1) with a dead-time correction process, as shown in Table~\ref{tab1}. The average value of the \emph{F} factors is $4.95\times10^{-7}$ counts/n with uncertainty of 0.7\%. The No. 2 \# sample was re-calibrated after two days from the last calibration, and the activity of residual activated atoms was deducted in the re-calculation of \emph{F} factor process. It can be seen that the activation diagnostic is stable from this two calibrations from sample 2\#.
\begin{center}
\tabcaption{ \label{tab1} The \emph{F} factors of each sample calibrated on accelerator neutron source. FWHM is the full width at half maximum of full-energy peak of interest, and \emph{dtr} is the dead-time ratio of germanium detector during sample measurement. Y$_c$ is the equivalent neutron yield at the end of irradiation by using Eq.~\ref{eq3}.}
\footnotesize
\begin{tabular}{cccccc}
\toprule[0.8pt]
\multirow{3}{1cm}{Sample No.} & \multicolumn{3}{c}{full-energy peak} & \multirow{3}{1cm}{Y$_c$ ($\times 10^{10}$)} & \multirow{3}{1.2cm}{\emph{F} factor ($\times 10^{-7}$ counts/n) } \\
\cline{2-4}
 &net & FWHM & \multirow{2}*{\emph{dtr} } &  \\
 & counts & /keV & \\
\hline
2$\#$ & 8344 & 1.05 & 0.83$\%$ & 8.63 & 4.84 \\
3$\#$ & 7539 & 1.05 & 0.74$\%$ & 7.52 & 5.01 \\
4$\#$ & 7984 & 1.29 & 0.93$\%$ & 8.09 & 4.94 \\
5$\#$ & 7733 & 1.05 & 0.89$\%$ & 8.05 & 4.81 \\
6$\#$ & 7338 & 1.06 & 0.79$\%$ & 7.07 & 5.19 \\
7$\#$ & 9492 & 1.06 & 0.84$\%$ & 9.39 & 5.06 \\
8$\#$ & 9341 & 1.04 & 0.85$\%$ & 9.54 & 4.90 \\
9$\#$ & 8434 & 1.01 & 0.82$\%$ & 8.60 & 4.90 \\
10$\#$ & 8231 & 1.05 & 0.78$\%$ & 8.24 & 4.99 \\
2$\#$ & 6889 & 1.04 & 0.72$\%$ & 7.12 & 4.85 \\

\bottomrule[0.8pt]
\end{tabular}
\end{center}

The indium samples are not only sensitive to DD neutrons but also slightly sensitive to scattered neutrons during calibration due to the lower threshold of 336 keV. The scattered neutron background comes mainly from the scattering of DD neutrons on the wall of accelerator room and the apparatus in this room, which can be measured by placing a neutron shield in front of the indium sample. The proper size of this shield was designed by a widely used Monte Carlo software named Geant 4 in version of 9.2.p04, as schematically shown in Fig.~\ref{fig6} (a). The neutrons transmitting this shield were simulated and their energy distribution is shown in Fig.~\ref{fig6} (b). The activity of sample attached this shield was reduced to 0.2\% of unshielded sample according to the simulation result. The primary neutrons entering the sample were very few by placing this shield in front of the indium sample, and the scattered neutron background from the accelerator room was measured. The measured activity of sample induced by scattered neutrons is $4.3\times10^{-8}$ counts/n with uncertainty of about 35\%. The big uncertainty mainly comes from the error of the shield aiming.
\begin{center}
\includegraphics[width=7cm]{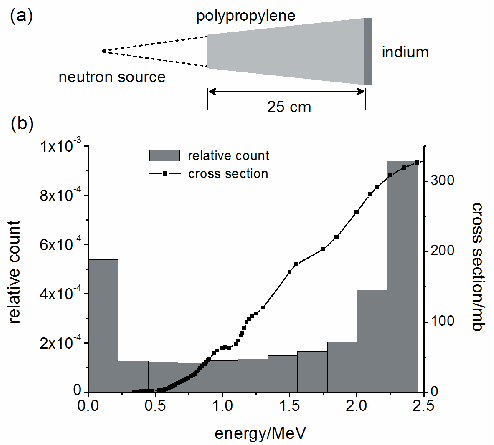}
\figcaption{\label{fig6}  (a) The schematic of scattered neutron measurement. The neutron shield is a conical block made of polypropylene with a 25 cm length. (b) The energy distribution of neutrons transmitting this shield simulated by Geant4.9.2.p04 codes (histogram). The dotted-black line is the inelastic scattering cross sections of $^{115}$In varying with neutron energies based on ENDF/B-VII.0 data.}
\end{center}

Deducting the neutron background, the \emph{F} factor induced by primary neutrons is $4.52\times10^{-7}$ counts/n with uncertainty of 3.4\%. There are some other error sources that affect calibration, such as the uncertainty caused by the energy difference of DD fusion neutrons from the accelerator and those from ICF (1\%), the uncertainty of neutron yield monitoring (1.6\%) \cite{lab13}, and the uncertainty of sample's solid angle (1.7\%). Thus it can be calculated that the total uncertainty of the \emph{F} factor is 4.3\% by the standard error synthesis formula. Adding the uncertainty of gamma net-counts and other errors during experimental measurement, the total uncertainty of neutron yield measurement can be as low as about 5\% for the implosion experiment with yield above $10^{12}$.

\section{Conclusion}
The indium activation diagnostic used as DD neutron yield measurement was calibrated on the accelerator neutron source located at Institute of Nuclear Physics and Chemistry, and the total uncertainty of the calibrated factor was 4.3\% based on the associated particle method. The scattered neutron background of the accelerator room was measured by placing a 25 cm thick polypropylene shield in front of indium sample, and the activity of sample induced by these scattered neutrons was deducted during calibration (about 9\% of the total). A decay curve analysis technique was used to discriminate gamma interferences by measuring the half life of activated sample, and the analysis result confirmed that the possible interference nuclides had a negligible effect on the calibrated factor in our samples. The linear range of activation diagnostic was measured, and it reached up to 100 cps net count rate within full energy peak of interest. When a 7 cm diameter by 1 cm thick sample placed at a 40 cm distance from target, the maximum neutron yield is about $5\times10^{12}$ measured by this activation diagnostic. By choosing appropriate sample size and distance from target, the higher DD neutron yields in the future implosion experiments can also be measured by this indium activation diagnostic with an accuracy of about 5\%.
\\

\acknowledgments{The authors are very grateful to professor Li Jiang, professor Bencao Lou and the staffs of the accelerator neutron source facility at Institute of Nuclear Physics and Chemistry for their cooperation.}

\end{multicols}

\vspace{-1mm}
\centerline{\rule{80mm}{0.1pt}}
\vspace{2mm}

\begin{multicols}{2}

\end{multicols}

\clearpage

\end{CJK*}
\end{document}